\documentclass[sigconf]{acmart}

\AtBeginDocument{%
  }

\copyrightyear{2025}
\acmYear{2025}
\setcopyright{cc}
\setcctype{by}
\acmConference[CIKM '25]{Proceedings of the 34th ACM International Conference on Information and Knowledge Management}{November 10--14, 2025}{Seoul, Republic of Korea}
\acmBooktitle{Proceedings of the 34th ACM International Conference on Information and Knowledge Management (CIKM '25), November 10--14, 2025, Seoul, Republic of Korea}
\acmDOI{10.1145/3746252.3760902}
\acmISBN{979-8-4007-2040-6/2025/11}

\usepackage{multirow} 
\usepackage{booktabs} 
\usepackage{caption} 
\usepackage{enumitem}
\definecolor{currentcolor}{RGB}{78,149,219}
\definecolor{nextcolor}{RGB}{164,116,186}
\begin{document}
\settopmatter{authorsperrow=4}

\newcommand{\method}{\textsc{Lincoln}}
\newcommand{\codelink}{https://github.com/ssong915/LINCOLN}

\title{Learning Short-Term and Long-Term Patterns of High-Order Dynamics in Real-World Networks}


\author{Yunyong Ko}
\affiliation{%
  \institution{Chung-Ang University}
  \city{Seoul}
  \country{Korea}
}
\email{yyko@cau.ac.kr}

\author{Da Eun Lee}
\affiliation{%
  \institution{Hanyang University}
  \city{Seoul}
  \country{Korea}
}
\affiliation{%
  \institution{KT Corporation}
  \city{Seoul}
  \country{Korea}
}
\email{ddanable.05@kt.com}

\author{Song Kyung Yu}
\affiliation{%
  \institution{Hanyang University}
  \city{Seoul}
  \country{Korea}
}
\affiliation{%
  \institution{KT Corporation}
  \city{Seoul}
  \country{Korea}
}
\email{songkyung.yu@kt.com}

\author{Sang-Wook Kim}
\authornote{Corresponding author}

\affiliation{%
  \institution{Hanyang University}
  \city{Seoul}
  \country{Korea}
}
\email{wook@hanyang.ac.kr}

\renewcommand{\shortauthors}{Yunyong Ko, Da Eun Lee, Song Kyung Yu, and Sang-Wook Kim}
\begin{abstract}
  Real-world networks have high-order relationships among objects and they evolve over time. 
  To capture such dynamics, many works have been studied in a range of fields.
  Via an in-depth preliminary analysis, we observe two important characteristics of high-order dynamics in real-world networks: high-order relations tend to \textbf{(O1)} \textit{have a structural and temporal influence on other relations in a short term} and \textbf{(O2)} \textit{periodically re-appear in a long term}.
  In this paper, we propose \textbf{{\method}}, a method for \textbf{\underline{L}}earning h\textbf{\underline{I}}gh-order dy\textbf{\underline{N}}ami\textbf{\underline{C}}s \textbf{\underline{O}}f rea\textbf{\underline{L}}-world \textbf{\underline{N}}etworks,
  that employs (1) bi-interactional hyperedge encoding for short-term patterns, (2) periodic time injection and (3) intermediate node representation for long-term patterns. 
  Via extensive experiments, we show that {\method} outperforms nine state-of-the-art methods in the dynamic hyperedge prediction task.
\end{abstract}



\maketitle

\section{Introduction}\label{sec-intro}

In real-world networks, high-order (i.e., \textit{group-wise}) relations are prevalent, 
such as (1) a paper co-authored by a group of researchers in collaboration networks and (2) a chemical reaction co-induced by a group of proteins in protein-protein interaction (PPI) networks. 
A \textit{hypergraph} can naturally model such a group-wise relation among an arbitrary number of objects as a \textit{hyperedge} without any information loss unlike an ordinary graph~\cite{ko2025enhancing,dong2020hnhn,feng2019hgnn}.
Due to its high expressive power, machine learning on hypergraphs has achieved breakthroughs in many tasks including node classification~\cite{feng2019hgnn,kang2024lowclass}, clustering~\cite{tsitsulin2023graph,jure-enron}, and link prediction~\cite{patil2020negative,chien2021allset,yu2025hygen}.

One important characteristic of real-world networks is that they are evolving \textit{dynamically}, 
where some objects and relations can (dis)appear over time.
For instance, a new researcher can join a research community (i.e., a new object appears), 
and a group of researchers may co-author a paper (i.e., a new relation is formed).

A number of dynamic graph learning methods~\cite{xu2020tgat,rossi2006tgn,you2022roland,he2023graphmixer,cong2023we} have been proposed to capture such dynamics.
However, they inevitably suffer from information loss in modeling original group-wise relations~\cite{feng2019hgnn,choe2023whatsnet}.
Although hypergraph learning have been widely studied to address this issue, 
most of existing methods have focused on \textit{static} networks where the structures are fixed~\cite{dong2020hnhn,yang2022hyperle,yadati2020nhp}.

\begin{figure}[t]
\centering
\begin{tabular}{cc}
    \centering
    \includegraphics[width=0.38\linewidth]{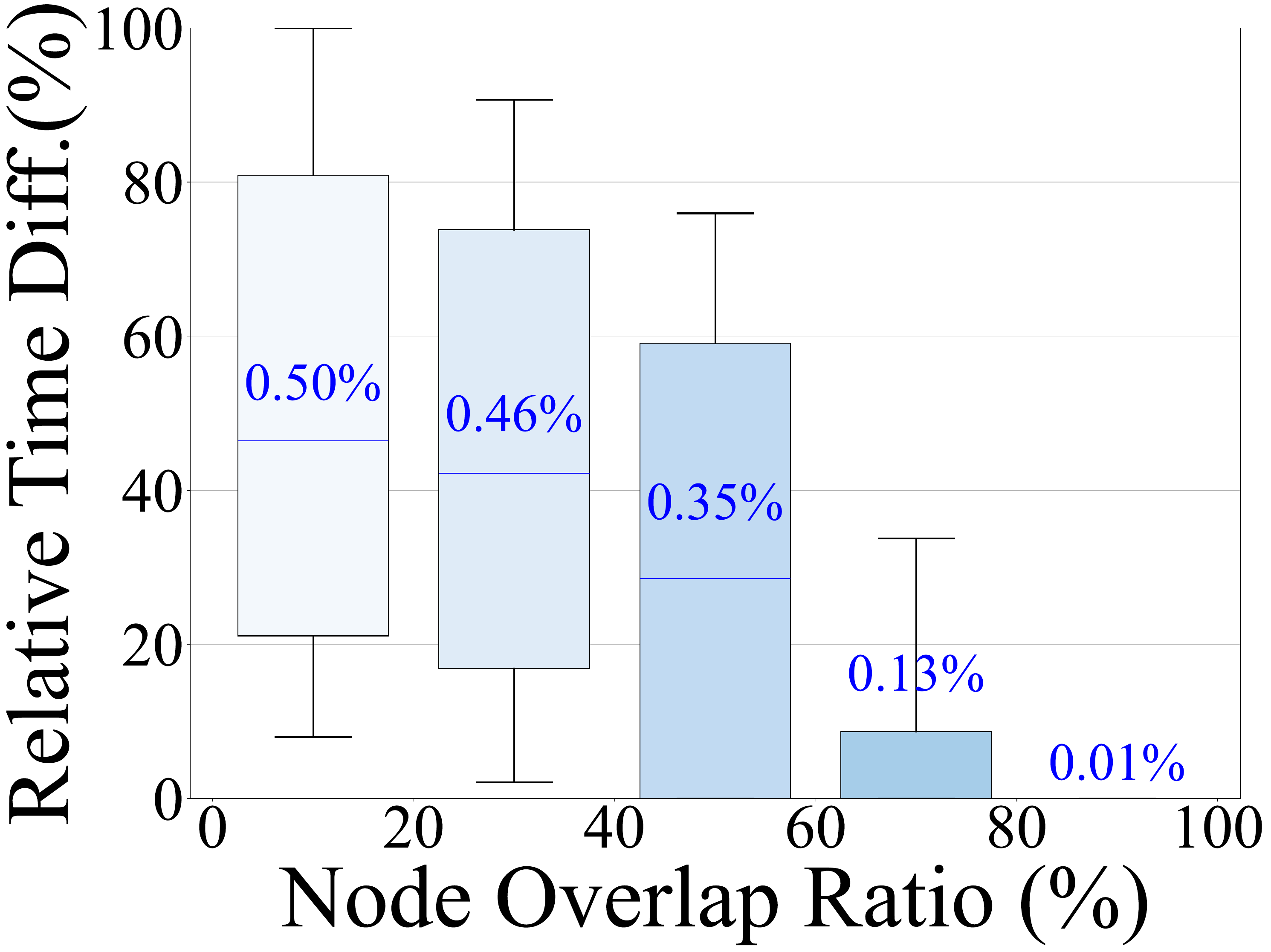} &
    \includegraphics[width=0.38\linewidth]{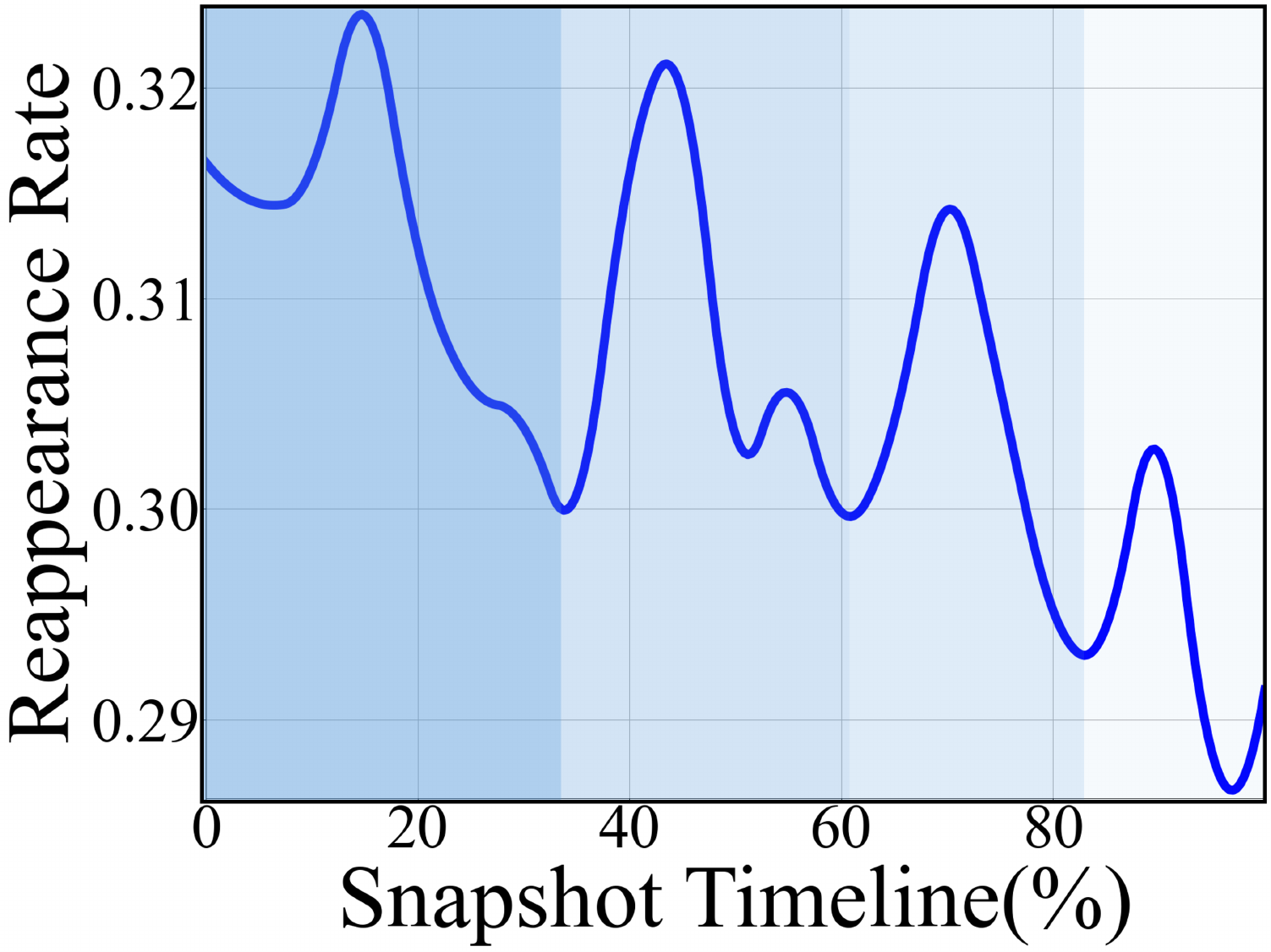} \\
    (a) Observation 1 (short-term) & (b) Observation 2 (long-term)
    \vspace{-3mm}
\end{tabular}
\caption{Observations: (a) short-term and (b) long-term patterns of high-order relations in real-world networks.}
\vspace{-5mm}
\label{fig:observations}
\end{figure}

From this motivation, in this paper, we propose a novel \textit{snapshot-based} hypergraph learning method, \textbf{{\method}}, for effectively capturing high-order dynamics of real-world networks. 
The process of {\method} is two-fold: 
given a dynamic network (i.e., a sequence of snapshots), (1) (\textbf{intra-snapshot learning}) it models each network snapshot as a hypergraph and represents the nodes in the hypergraph as embeddings by a hypergraph encoder and (2) (\textbf{inter-snapshot learning}) it updates the node embeddings based on the nodes and hyperedges newly observed in the next snapshot to capture temporal patterns of the evolving network structure.

For better understanding high-order dynamics, 
we randomly selected 50 high-order relations from the first 20\% snapshots of the Congress dataset~\cite{benson2018dataset} and investigated the relationships among the selected relations within each snapshot (short-term) and how many times they \textit{re-appear} in the next snapshots (long-term).

\vspace{1mm}
\noindent
\textbf{(O1) Short-term pattern.} 
Figure~\ref{fig:observations}(a) shows the average difference between the formation times of a pair of hyperedges according to their node overlap (i.e., structural similarity) within each snapshot.
The result shows that \textit{the more structurally-similar high-order relations tend to be formed in shorter time intervals}.
which implies that `high-order relations within a snapshot have a structural and temporal influence on the formation of future high-order relations'.
Thus, it is crucial to reflect the structural and temporal influence on the formation of high-order relations in the intra-snapshot learning.

\vspace{1mm}
\noindent
\textbf{(O2) Long-term pattern.} 
Figure~\ref{fig:observations}(b) shows the re-appearance rate of high-order relations across snapshots, 
where different background colors represent different snapshots.
Clearly, \textit{high-order relations tend to periodically re-appear}, 
which indicates that `high-order relations across snapshots are repeated in a long-term interval'.
Therefore, it is important to consider the long-term periodic pattern of high-order relations in the inter-snapshot learning.

Based on these observations, we propose three strategies: 
\textbf{(1) Bi-interactional hyperedge encoding} that captures structural and temporal patterns of high-order relations within a snapshot,
\textbf{(2) Periodic time injection} that reflects long-term periodic time information into node embeddings,
and \textbf{(3) intermediate node representation} that learns the temporal patterns of intermediate node embeddings across snapshots.

The main contributions of this work are as follows.
\vspace{-1mm}
\begin{itemize}[leftmargin=10pt]
    \item \textbf{Observations}: We observe that high-order relations tend to \textbf{(O1)} \textit{have a structural and temporal influence on other relations in a short term} and \textbf{(O2)} \textit{periodically re-appear in a long term}.
    \item \textbf{Method}: We propose a novel dynamic hypergraph learning method, {\method}, that effectively captures long-term and short-term patterns of high-order relations in real-world networks.
    \item \textbf{Evaluation}: Via extensive experiments on seven real-world datasets, we demonstrate that {\method} outperforms nine state-of-the-art methods in the dynamic hyperedge prediction task.
\end{itemize}

\vspace{-1mm}
\noindent
For reproducibility, we have released the code of {\method} and datasets at: \url{\codelink}.

\vspace{-2mm}
\section{Related Work}\label{sec-related}

\noindent
\textbf{Graph learning methods.}
A number of graph learning methods have been proposed.
TGAT~\cite{xu2020tgat} considers the time difference between relations in learning temporal node embeddings.
TGN~\cite{rossi2006tgn} learns temporal node embeddings based on nodes’ neighbors and their long-term dependencies.
ROLAND~\cite{you2022roland} extends static GNNs to a dynamic setting, 
while utilizing hierarchical node embeddings for better node representations.
These methods, however, inevitably suffer from information loss in modeling high-order relations.

\vspace{1mm}
\noindent
\textbf{Hypergraph learning methods.}
To address the information loss, 
hypergraph learning methods~\cite{ko2025enhancing,feng2019hgnn,dong2020hnhn,hwang2022ahp,choe2023whatsnet}, modeling a network as a \textit{hypergraph}, have been widely studied. 
HGNN~\cite{feng2019hgnn} learns high-order relations based on a clique expansion, replacing hyperedges with cliques to transform a hypergraph into a graph. 
HNHN~\cite{dong2020hnhn} represents a hypergraph as a bipartite graph and learns both node and hyperedge embeddings. 
AHP~\cite{hwang2022ahp} employs an adversarial training-based model to generate realistic negative hyperedges.
WHATsNET~\cite{choe2023whatsnet} learns within-hyperedge relations among nodes via attention and positional encoding schemes for the edge-dependent node classification.
Although these methods often outperform graph learning methods in many downstream tasks, 
they focus mainly on \textit{static} networks where the structures are fixed. 
There have been only a handful of dynamic hypergraph learning methods~\cite{jiang2019dhgnn,zhou2023tdhnn,liu2022hit}.
DHGNN~\cite{jiang2019dhgnn} and TDHNN~\cite{zhou2023tdhnn}, define hyperedges and re-construct a hypergraph structure based on the defined hyperedges. 
These methods, however, often fail to capture temporal patterns of the network structure.

\begin{figure}[t]
\centering
\includegraphics[width=1.0\linewidth]{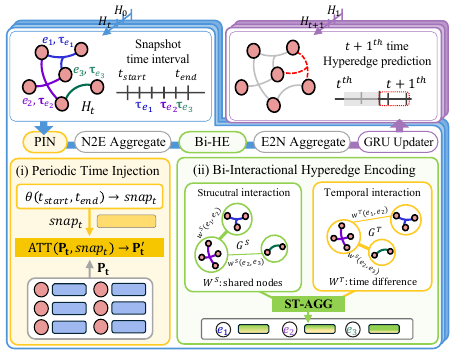}
\vspace{-5mm}
\caption{Overview: (1) \textcolor{currentcolor}{intra-snapshot learning} for the structural features within a snapshot and (2) \textcolor{nextcolor}{inter-snapshot learning} for the temporal features across snapshots.}
\vspace{-3mm}
\label{fig:overview}
\end{figure}

\section{Proposed Method: {\method}}\label{sec-proposed}

\noindent
\textbf{Dynamic hypergraph}.  
Consider a dynamic hypergraph as a sequence of hypergraph snapshots $H_D=\{H_1, H_2, \dots, H_T\}$, 
where each snapshot $H_t = (V_t, E_t)$ consists of a set of nodes $V_t=\{v_{(1,t)},$ $v_{(2,t)}$ $,..., v_{{(|V_t|,t)}}\}$ and a set of hyperedges $E_t=\{e_{(1,t)}, e_{(2,t)},$ $..., e_{(|E_t|,t)}\}$.
A hypergraph snapshot is represented by an \textit{incidence} matrix $\mathbf{H_t}\in \{0,1\}^{|V_t|\times |E_t|}$,
where each element $h_{ij}=1$ if $v_i \in e_j$, and $h_{ij}=0$ otherwise.
In each snapshot, node and hyperedge features are denoted as $\mathbf{P}_t \in \mathbb{R}^{|V_t| \times d}$ and $\mathbf{Q}_t \in \mathbb{R}^{|E_t| \times d}$, respectively.

\vspace{1mm}
\noindent
\textbf{\textsc{Problem 1} (\textsc{Dynamic Hyperedge Prediction}).} 
Given a dynamic hypergraph $H_D=\{H_1, H_2, ..., H_T\}$ and the initial node features $\mathbf{X}\in \mathbb{R}^{|V|\times F}$,
we aim to predict whether a hyperedge candidate $e'$ belongs to $E_{t+1}$ based on the previous $t$ snapshots.


\vspace{2mm}
\noindent
\textbf{Overview of {\method}.}
As illustrated in Figure~\ref{fig:overview}, 
{\method} aims to learn (1) the structural and temporal features within a snapshot (i.e., \textit{intra-snapshot learning}) and 
(2) the evolving networks across snapshots (i.e., \textit{inter-snapshot learning}).

\vspace{1mm}
\noindent
\textbf{(1) Intra-Snapshot Learning.}
Given a hypergraph snapshot $H_t = (V_t, E_t)$, 
{\method} produces node representations $\mathbf{P}_t \in \mathbb{R}^{|V_t| \times d}$ and hyperedge representations $\mathbf{Q}_t \in \mathbb{R}^{|E_t| \times d}$ by using a hyperedge encoder.
Following~\cite{dong2020hnhn,chien2021allset,ko2025enhancing}, {\method} adopts a \textit{2-stage aggregation} approach,
which repeats (1) (\textit{node-to-hyperedge}) producing the hyperedge embeddings by aggregating the node embeddings and (2) (\textit{hyperedge-to-node}) producing the node embeddings by aggregating the hyperedge embeddings:
\begin{align}
    \mathbf{Q_t}^{(k)} &=  \sigma(\mathbf{D}^{-1}_{E_t}\mathbf{H_t}^{T}\mathbf{P_t}^{(k-1)} \mathbf{W}^{(k)}_{E_t} + b^{(k)}_{E_t})\label{eq:node-to-edge}, \\
    \mathbf{P_t}^{(k)} &= \sigma(\mathbf{D}^{-1}_{V_t}\mathbf{H_t}\mathbf{Q_t}^{(k)} \mathbf{W}^{(k)}_{V_t} + b^{(k)}_{V_t}),\label{eq:edge-to-node}
\end{align}
where 
$\mathbf{P_t}^{(k)}$ and $\mathbf{Q_t}^{(k)}$ are the node and hyperedge embeddings at the \textit{k}-th layer,
$\mathbf{W}^{(k)}_*$ and $b^{(k)}_*$ are trainable weight and bias matrices, respectively, 
$\mathbf{D}^{-1}_*$ is a normalization term, and $\sigma$ is a non-linear activation function.

\vspace{1mm}
\noindent
\textbf{\underline{(1)-(a) Periodic time injection}}: 
Given the node embeddings $\mathbf{P_t}$ at time $t$, {\method} injects \textit{period-specific} time information into the embeddings of nodes for reflecting the periodic patterns of high-order relations into them, before the N2E aggregation.
Specifically, it (i) produces a snapshot-time embedding $snap_t$ by passing the start and end times of a snapshot, $t_{start}$ and $t_{end}$, into a periodic function $\theta(\cdot)$ and (ii) aggregates $snap_t$ with $\mathbf{P_t}$ based on their attention weights computed by the snapshot-aware attention mechanism, to reflect different importance of each node in the context of the snapshot:    
\begin{align}
    snap_t = \theta(t_{start}, t_{end}), \hspace{2mm} \mathbf{P'_t} = \textsc{Att}(\mathbf{P_t}, snap_t).
\end{align}
For a periodic function $\theta(\cdot)$, we use a cosine function.    
The resulting snapshot-aware node embeddings ($\mathbf{P'_t}$) are used in the N2E aggregation (Eq.\ref{eq:node-to-edge}),
enabling {\method} to incorporate periodic patterns of high-order relations into its learning process.

\vspace{1mm}
\noindent
\textbf{\underline{(1)-(b) Bi-interactional hyperedge encoding}}: 
From the observation \textbf{(O1)}, the high-order relations within a snapshot play a structural and temporal role in forming other relations.
To capture such an important property,
{\method} (i) disentangles a high-order relation into structural and temporal embeddings and (ii) learns structural and temporal correlations among high-order relations.
Specifically, {\method} constructs two graphs (See $G^S$ and $G^T$ in Figure~\ref{fig:overview}), 
where each node corresponds to a hyperedge and two hyperedge nodes are connected if they have common nodes.
The hyperedge nodes in each graph are initialized with $\mathbf{Q_t}$ (Eq.~\ref{eq:node-to-edge}). 
We assign edge weights $w^S(e_i,e_j)$ and $w^T(e_i,e_j)$ in $G^S$ and $G^T$ based on the following intuitions:
two hyperedges are likely to be closely connected more if they share more nodes (i.e., high structural proximity) or they occur in a shorter time period (i.e., high temporal proximity).
Then, each hyperedge embedding is represented into structural and temporal embeddings by applying a GNN model to $G^S$ and $G^T$, respectively:
\begin{align}
    \mathbf{Q^S_t} = \textsc{Gnn}^S(\mathbf{Q_t}, G^S),\hspace{0.8em} \mathbf{Q^T_t} = \textsc{Gnn}^T(\mathbf{Q_t}, G^T).
\end{align}
For a GNN model, we use GCN~\cite{kipf2016gcn}.
The two embeddings for each hyperedge are combined by an aggregation function:
\begin{equation}
    \mathbf{Q'_t} = \textsc{ST-Agg}(\mathbf{Q^S_t}, \mathbf{Q^T_t}) = \sigma((\mathbf{Q^S_t} \oplus \mathbf{Q^T_t})\cdot \mathbf{W}_{ST} + \mathbf{b}_{ST}),    
\end{equation}
where $\mathbf{W}_{ST}$ and $\mathbf{b}_{ST}$ are the trainable parameters.
The resulting \textit{structural-and-temporal} hyperedge embeddings ($\mathbf{Q'_t}$) are used in the E2N aggregation (Eq.~\ref{eq:edge-to-node}),
thereby reflecting the structural and temporal patterns of high-order relations into the learning process.

\vspace{1mm}
\noindent
\textbf{(2) Inter-Snapshot Learning.}
Then, {\method} aims to learn their temporal patterns across snapshots.
Formally, {\method} updates the node embeddings at time $t$ ($\mathbf{P_t}$) 
based on the node embeddings from the previous snapshot ($\mathbf{P^*_{t-1}}$) by using a temporal feature update module $g(\cdot)$:
\begin{equation}
    \mathbf{P^{*}_t} = g(\mathbf{P_t}, \mathbf{P^{*}_{t-1}}).
\end{equation}
For a temporal feature update module, we use GRU~\cite{cho2014gru}.

\vspace{1mm}
\noindent
\textbf{\underline{(2)-(a) Intermediate node representation}}: 
Most existing dynamic learning methods~\cite{yin2022dynamic,pareja2020evolvegcn,zhao2019t} use the node embeddings from only the \textit{final} layer of intra-snapshot learning (i.e., $\mathbf{P^{(k)}_t}$) as the input of $g(\cdot)$.
However, the node embeddings generated in the intermediate steps of intra-snapshot learning have useful information, 
which is hidden across a sequence of snapshots and beneficial to capturing temporal patterns of evolving networks over time~\cite{you2022roland}.
Thus, we additionally leverage the intermediate node embeddings (i.e., $\mathbf{P^{(l)}_t}, 0\leq l\leq k$) as the input of $g(\cdot)$ and learn their temporal patterns for better understanding the complex network structure:
\begin{equation}
    \mathbf{P^{*(l)}_t} = g(\mathbf{P^{(l)}_t}, \mathbf{P^{*(l)}_{t-1}}).
\end{equation}

\vspace{1mm}
\noindent
\textbf{(3) Prediction and Training}.
Given the final node embeddings $\mathbf{P^*_t}$ at time $t$ and a hyperedge candidate $e'$ at time $t+1$,
{\method} aggregates the embeddings of the nodes in $e'$, $\mathbf{P_t}^*[e',:]$
and computes the probability of $e'$ belonging to $E_{t+1}$, $\hat{y}_{e'}$, as: 
\begin{align}
    \hat{y}_{e'} = pred(q^*_{e'}), \hspace{2mm} q^*_{e'} = \textsc{Agg}(\mathbf{P_t}^*[e',:]),
    \label{eq:prediction}
\end{align}
where $q^*_{e'} \in \mathbb{R}^{d}$ is the final hyperedge candidate embedding,
$pred(\cdot)$ is a hyperedge predictor, a fully-connected layer ($d \times 1$) followed by a sigmoid function,
and $\textsc{Agg}(\cdot)$ is an element-wise average pooling.

For training, 
we consider positive and negative hyperedges.
For negative hyperedges, we use the motif negative sampling method (MNS)~\cite{patil2020negative}.
Thus, {\method} aims to train its model parameters so that positive examples obtain higher scores while negative examples obtain lower scores.
Formally, the prediction loss is defined as:
\begin{align}
    \mathcal{L}_{pred}=-\frac{1}{|E'_{t+1}|} \sum_{e'\in E'_{t+1}} y_{e'}\cdot\log{\hat{y}_{e'}}+(1-y_{e'})\cdot\log{(1-\hat{y}_{e'})} \label{eq:bceloss}
\end{align}
where $y_{e'}$ is the label of the hyperedge candidate $e'$ (1 or 0).

\vspace{1mm}
\noindent
\textbf{\underline{(3)-(a) Contrastive learning}}: 
In addition, 
we use contrastive loss $\mathcal{L}_{con}$ between the structural and temporal hyperedge embeddings for better learning high-order relations, which is defined as:
\begin{align}
    \mathcal{L}_{con} = -\log{sim(\mathbf{Q^S_t}, \mathbf{Q^T_t})},
    \label{eq:loss-contrastive}
\end{align}
where $sim(\cdot)$ is the cosine similarity.
Finally, the two losses are unified with a hyperparameter $\beta$:
\begin{align}
    \mathcal{L} = \mathcal{L}_{pred} + \beta\mathcal{L}_{con}.
    \label{eq:loss-total}
\end{align}

\begin{table*}[t]
\small
\centering
\footnotesize
\caption{Hyperedge prediction accuracy (\%) on seven dynamic hypergraphs. For each dataset, the best and the second-best results are highlighted in the \textbf{bold} font and \underline{underlined}, respectively. OOM indicates `out of memory' on a 11GB GPU.}\label{table:eval-accuracy}
\vspace{-3mm}
\setlength\tabcolsep{5.3pt} 
\def\arraystretch{0.7} 
\resizebox{\textwidth}{!}{
\footnotesize
\begin{tabular}{c|ccccccccccccccc}
\toprule
\multirow{2}{*}{Method} & \multicolumn{2}{c}{Email-enron}  & \multicolumn{2}{c}{Email-eu} & \multicolumn{2}{c}{Tags} & \multicolumn{2}{c}{Thread} & \multicolumn{2}{c}{Contact-primary} & \multicolumn{2}{c}{Contact-high} & \multicolumn{2}{c}{Congress} & \multirow{2}{*}{\textbf{Rank}}\\

\cmidrule(lr){2-3} \cmidrule(lr){4-5} \cmidrule(lr){6-7} \cmidrule(lr){8-9} \cmidrule(lr){10-11} \cmidrule(lr){12-13} \cmidrule(lr){14-15}

& {\scriptsize AUROC} & {\scriptsize AP} & {\scriptsize AUROC} & {\scriptsize AP} & {\scriptsize AUROC} & {\scriptsize AP} & {\scriptsize AUROC} & {\scriptsize AP} & {\scriptsize AUROC} & {\scriptsize AP} & {\scriptsize AUROC} & {\scriptsize AP} & {\scriptsize AUROC} & {\scriptsize AP} & \\

\midrule
 TGN     & 87.75 & 87.61 & 59.07 & 58.96 & 61.90 & 60.74 & \underline{84.43} & 82.25 & 58.30 & 58.74 & 75.19 & 71.65 & OOM & OOM & 5.1\\
 TGAT    & \textbf{90.93} & \underline{89.72} & 44.18 & 47.10 & 48.49 & 50.97 & 83.91 & 80.12 & 44.10 & 50.09 & 41.79 & 49.26 & OOM & OOM & 7.0\\
 ROLAND  & 80.37 & 82.42 & \underline{64.20} & \underline{66.12} & 67.37 & 67.04 & 80.76 & \underline{85.23} & \underline{74.47} & \underline{76.60} & 70.96 & 75.59 & \underline{94.01} & \underline{92.90} & \underline{3.0}\\

\midrule
 HGNN    & 64.34 & 67.54 & 50.17 & 52.05 & 54.26 & 54.85 & OOM & OOM & 62.15 & 60.33 & 58.03 & 58.11 & 53.87 & 53.06 & 7.6\\
 HNHN    & 74.29 & 70.79 & 55.55 & 53.78 & 57.53 & 55.56 & 50.48 & 50.68 & 56.44 & 54.77 & 59.04 & 59.21 & 58.57 & 57.10 & 7.2\\
 AHP     & 81.99 & 81.38 & 61.97 & 59.73 & 62.89 & 62.85 & 75.23 & 74.54 & 72.55 & 68.22 & \underline{83.68} & \underline{80.27} & 68.06 & 75.71 & 4.0\\
 WHATsNet& 76.79 & 79.29 & 60.25 & 59.98 & 63.63 & 62.36 & 51.12 & 53.06 & 59.19 & 59.57 & 57.24 & 61.03 & 46.84 & 53.02 & 6.1\\

\midrule
 DHGNN   & 74.86 & 75.14 & OOM & OOM & OOM & OOM & OOM & OOM & OOM & OOM & OOM & OOM & OOM & OOM & 9.3\\
 TDHNN   & 81.69 & 84.47 & 63.72 & 63.89 & \textbf{76.65} & \underline{78.63} & 65.22 & 67.64 & 71.14 & 71.03 & 73.36 & 74.58 & 72.08 & 72.22 & 3.6\\
\midrule

\textbf{} \textbf{{\method}} & \underline{84.35} & \textbf{92.90} &\textbf{66.16} & \textbf{77.85} & \underline{72.19} & \textbf{80.55} & \textbf{85.21}& \textbf{90.77} & \textbf{78.13} & \textbf{84.97} & \textbf{85.92} & \textbf{97.73} & \textbf{94.17} & \textbf{97.54} & \textbf{1.2} \\

\bottomrule
\end{tabular}
}
\end{table*}

\begin{table*}[t]
\centering

\caption{Ablation study: each of our proposed strategies is beneficial to improving the accuracy of {\method}.}\label{table:ablation-hne}
\vspace{-4mm}
\scriptsize
\setlength\tabcolsep{3pt} 
\def\arraystretch{0.5} 
\resizebox{\textwidth}{!}{
\begin{tabular}{c|cccccccccccccc}
\toprule
\multirow{2}{*}{Method} & \multicolumn{2}{c}{Email-enron}  & \multicolumn{2}{c}{Email-eu} & \multicolumn{2}{c}{Tags} & \multicolumn{2}{c}{Thread} & \multicolumn{2}{c}{Contact-primary} & \multicolumn{2}{c}{Contact-high} & \multicolumn{2}{c}{Congress} \\

\cmidrule(lr){2-3} \cmidrule(lr){4-5} \cmidrule(lr){6-7} \cmidrule(lr){8-9} \cmidrule(lr){10-11} \cmidrule(lr){12-13} \cmidrule(lr){14-15}

  & {\scriptsize AUROC} & {\scriptsize AP} & {\scriptsize AUROC} & {\scriptsize AP} & {\scriptsize AUROC} & {\scriptsize AP} & {\scriptsize AUROC} & {\scriptsize AP} & {\scriptsize AUROC} & {\scriptsize AP} & {\scriptsize AUROC} & {\scriptsize AP} & {\scriptsize AUROC} & {\scriptsize AP}\\
 \midrule
\textbf{{\method}} & \textbf{84.35} & \textbf{92.90} & \textbf{66.16} & \textbf{77.85} & \underline{72.19} & \underline{80.55} & \textbf{85.21} & 90.77 & \textbf{77.84} & \textbf{84.32} & \textbf{85.92} & \textbf{97.73} & \textbf{94.17} & \textbf{97.54} \\


\midrule

\textbf{{\method}} w/o Bi-HE & 72.01$\textcolor{blue}\downarrow$ & 86.54 $\textcolor{blue}\downarrow$ & \underline{65.13} $\textcolor{blue}\downarrow$ & \underline{77.00} $\textcolor{blue}\downarrow$ & \textbf{72.74} $\textcolor{red}\uparrow$ & \textbf{81.31} $\textcolor{red}\uparrow$ & \underline{81.78} $\textcolor{blue}\downarrow$& \textbf{94.47} $\textcolor{blue}\downarrow$ & 72.24$\textcolor{blue}\downarrow$ & 81.63$\textcolor{blue}\downarrow$ & 78.63$\textcolor{blue}\downarrow$ & 96.62$\textcolor{blue}\downarrow$ & 81.60$\textcolor{blue}\downarrow$ & 93.18$\textcolor{blue}\downarrow$ \\

\midrule

\textbf{{\method}} w/o PIN & \underline{75.84} $\textcolor{blue}\downarrow$ & \underline{88.60} $\textcolor{blue}\downarrow$ & 59.95$\textcolor{blue}\downarrow$ & 70.69 $\textcolor{blue}\downarrow$ & 70.82 $\textcolor{blue}\downarrow$ & 78.99 $\textcolor{blue}\downarrow$ & 80.03 $\textcolor{blue}\downarrow$ & \underline{94.17} $\textcolor{red}\uparrow$ & \underline{75.99} $\textcolor{blue}\downarrow$& \underline{83.91} $\textcolor{blue}\downarrow$ & \underline{82.23} $\textcolor{blue}\downarrow$ & \underline{96.90} $\textcolor{blue}\downarrow$ & \underline{92.03} $\textcolor{blue}\downarrow$& \underline{96.90} $\textcolor{blue}\downarrow$\\


\bottomrule
\end{tabular}
}
\end{table*}

\section{Experimental Validation}\label{sec-eval}

We evaluate {\method} by answering the following questions.
\begin{itemize}[leftmargin=10pt]
    \item \textbf{EQ1}. To what extent does {\method} improve the existing methods in terms of the hyperedge prediction on dynamic networks?
    \item \textbf{EQ2}. Is each of our strategies beneficial to learning high-order dynamics of real-world relations?    

    \item \textbf{EQ3}. Is the intermediate node representation effective in capturing temporal patterns of evolving networks?
\end{itemize}

\subsection{Experimental Setup} 

\noindent
\textbf{Datasets and competitors.}
We use seven real-world hypergraphs \newline~\cite{benson2018dataset}, which are classified into five categories:
(1) email, (2) tagging, (3) thread participant, (4) contact, and (5) congress networks.
We select 9 competitors, including 3 \textit{dynamic graph-based} methods~\cite{rossi2006tgn,xu2020tgat,you2022roland}, 4 \textit{static hypergraph-based} methods~\cite{feng2019hgnn,dong2020hnhn,hwang2022ahp,choe2023whatsnet}, and 2 \textit{dynamic hypergraph-based} methods~\cite{jiang2019dhgnn,zhou2023tdhnn}.

\vspace{1mm}
\noindent
\textbf{Evaluation protocol.}
We consider the \textit{hyperedge prediction} in a dynamic network.
We use a `live-update' evaluation setting, following ~\cite{you2022roland}.
It evaluates a model by utilizing \textit{all} snapshots in a uniform way.
It constructs training (70\%), validation (20\%), and test (10\%) sets by splitting hyperedges `within' each snapshot.
As metrics, we use \textit{area under ROC} (AUROC), and \textit{average precision} (AP).
We (1) measure AUROC on the test set at the epoch when the AUROC over the validation set is maximized, 
and (2) report the averaged AUROC and AP on the test set over five runs.

\subsection{Experimental Results} 

\begin{figure}[t]
\begin{tabular}{cc}
    \includegraphics[width=0.71\linewidth]{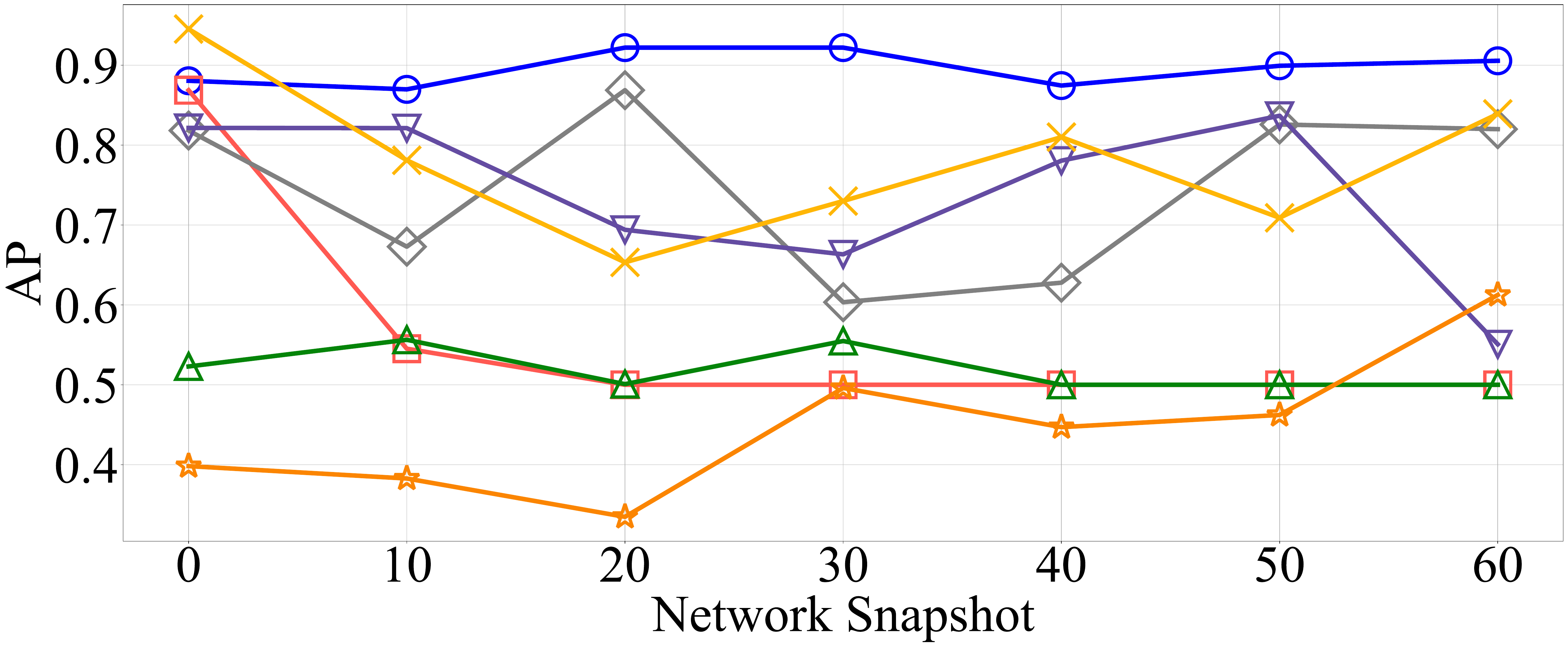} &
    \includegraphics[width=0.185\linewidth]{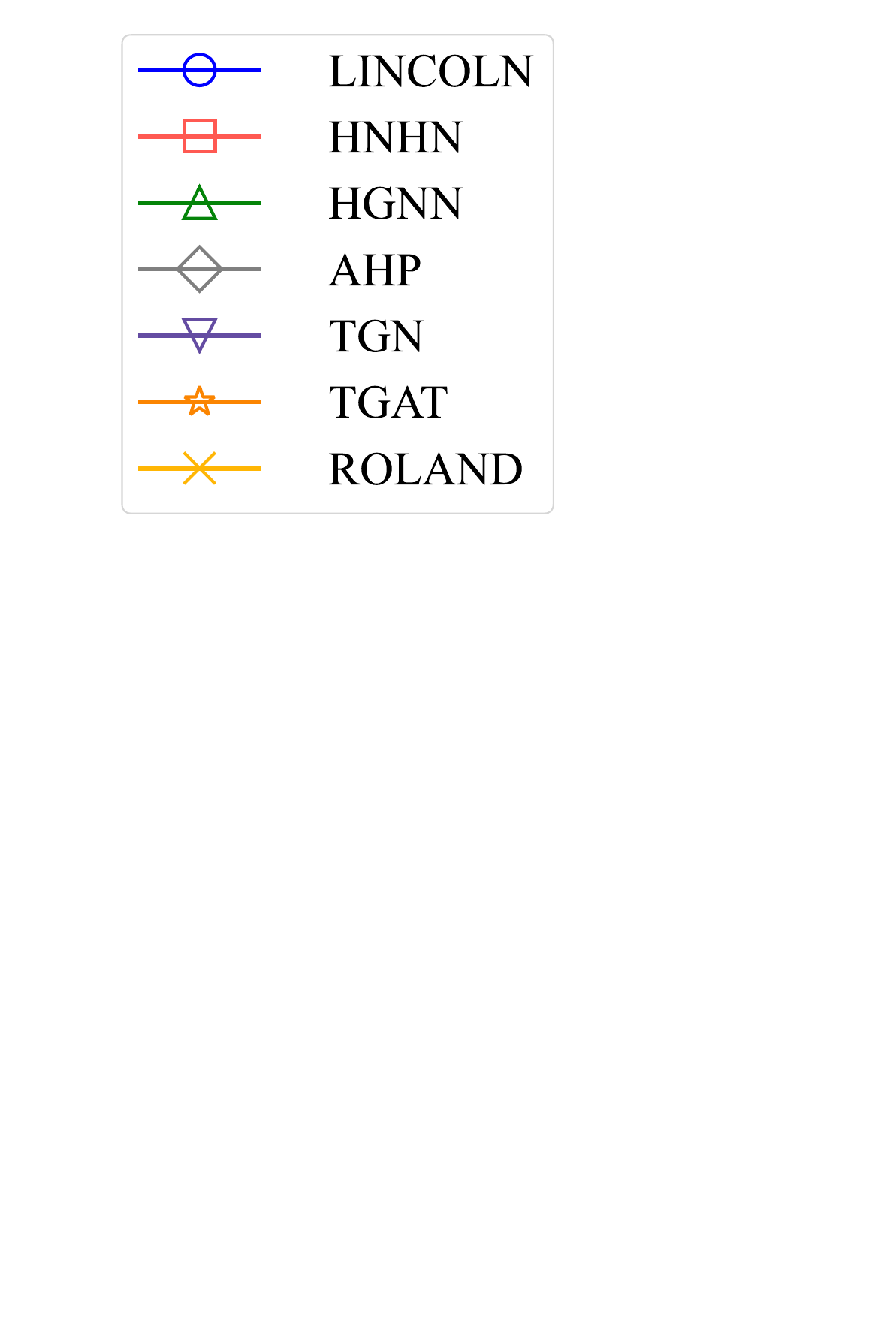} \\
\end{tabular}
\vspace{-4mm}
\caption{Hyperedge prediction accuracy across snapshots.}
\vspace{-5mm}
\label{fig:qualitative-analysis}
\end{figure}

\noindent
\textbf{Q1. Hyperedge prediction on dynamic networks.}
Table~\ref{table:eval-accuracy} shows the hyperedge prediction accuracy on seven dynamic hypergraphs.
{\method} significantly outperforms competing methods in most cases, while achieving the best-averaged rank.
Figure~\ref{fig:qualitative-analysis} shows the accuracy (AP) of each method at every 10 snapshots in the contact-primary dataset.
{\method} achieves the highest accuracies in most snapshots.
Moreover, the accuracy of {\method} on each snapshot tends to be \textit{stable}, while those of other methods tend to \textit{fluctuate steeply}.
These results imply that {\method} is able to effectively capture both short-term and long-term patterns of high-order dynamics better than other competing methods.

\vspace{1mm}
\noindent
\textbf{Q2. Ablation study.}
We verify the effectiveness of the following two strategies by ablating one of them: 
(a) bi-interactional hyperedge encoding and (b) periodic time injection.
Table~\ref{table:ablation-hne} shows that the original version of {\method} achieves the highest accuracy in most cases, which indicates that ablating one of our proposed strategies could lead to performance degradation.
Therefore, these results verify (1) the usefulness of our observations -- (O1) the structural and temporal dependencies among high-order relations in a short term and (O2) their periodic re-appearance in a long term -- and (2) the effectiveness of our proposed strategies for capturing the short-term and long-term high-order dynamics.

\begin{figure}[t]
\centering
\begin{tabular}{cc}
    \centering
    \includegraphics[width=0.435\linewidth]{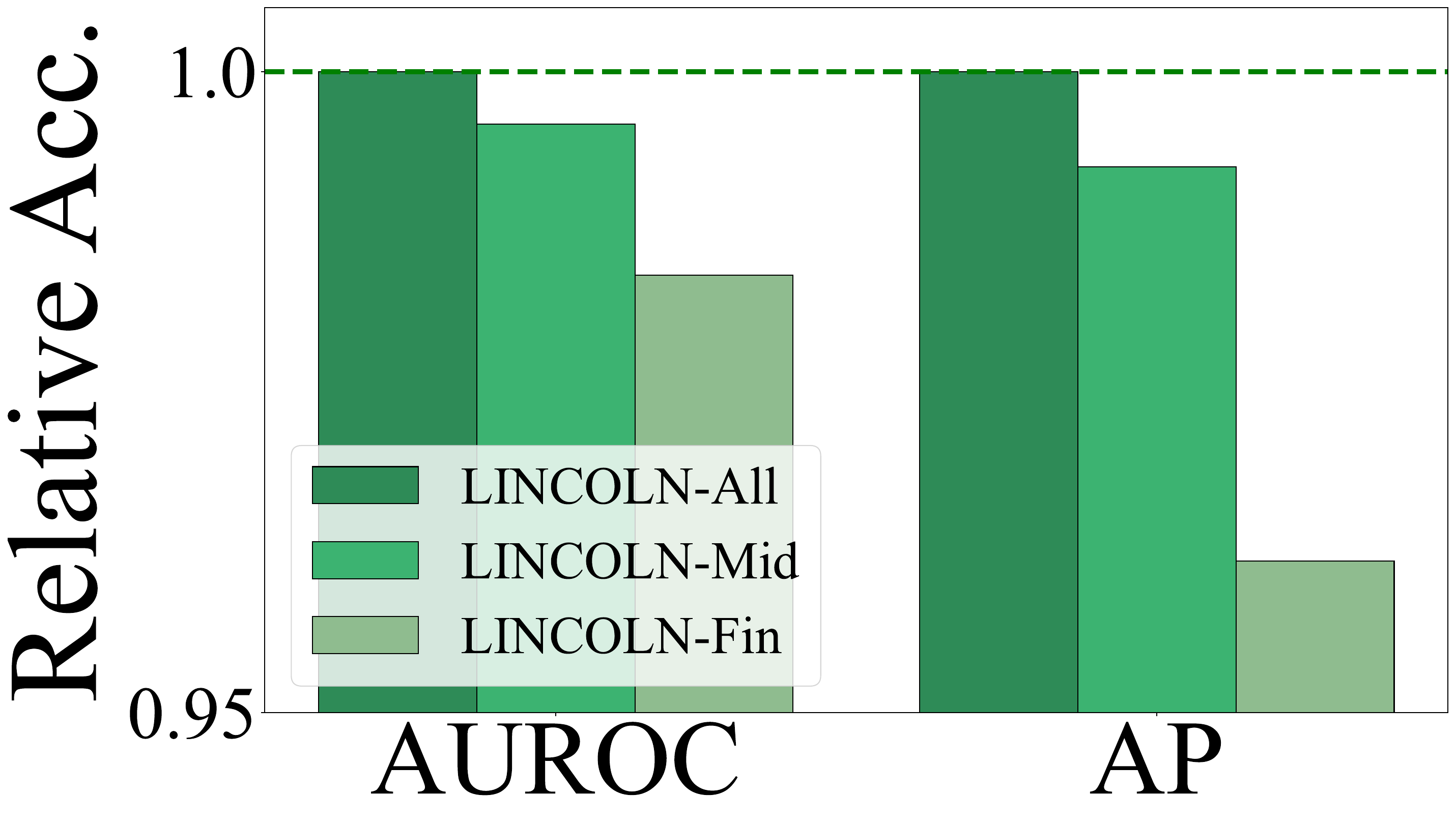} &
    \includegraphics[width=0.435\linewidth]{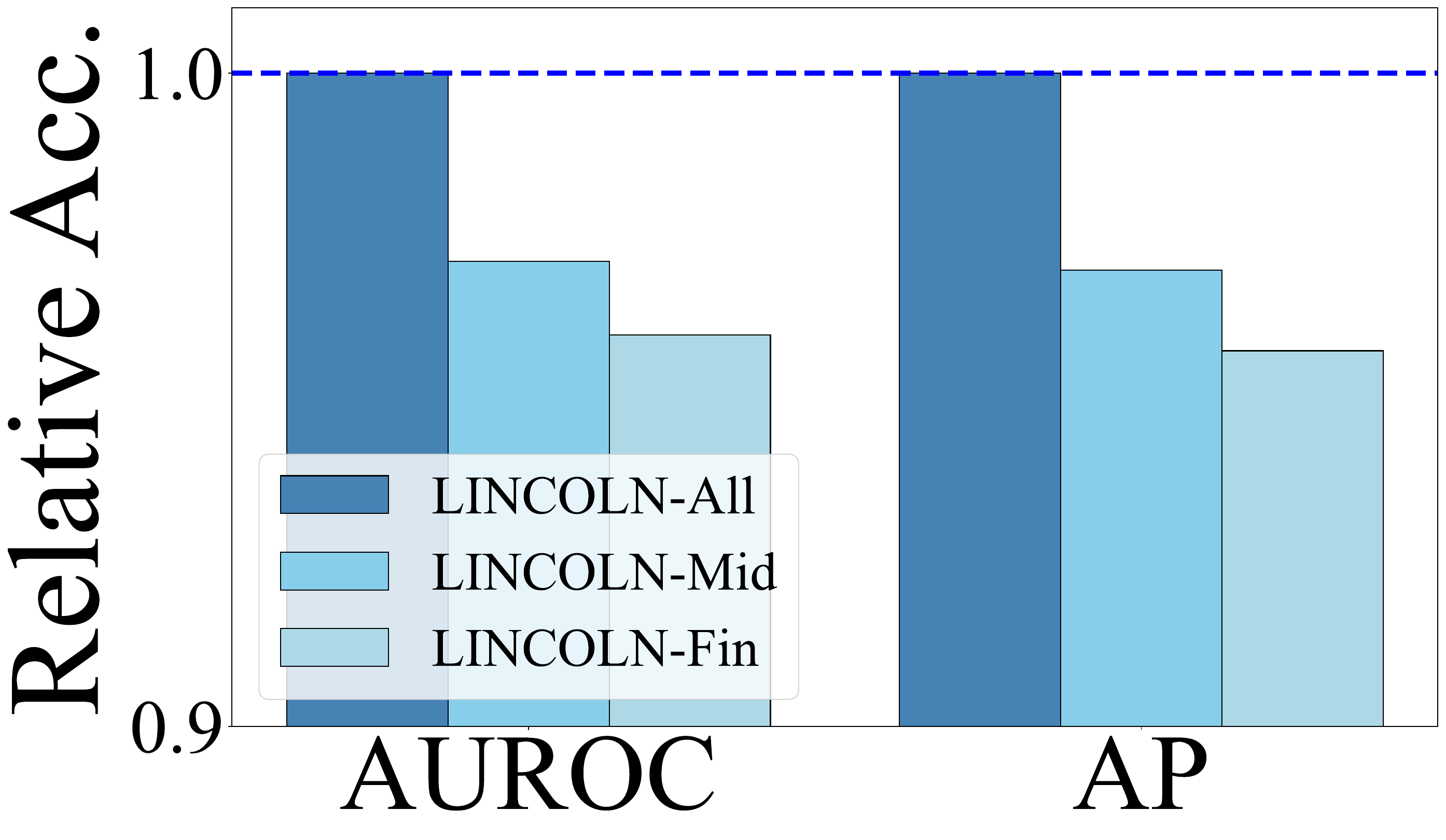} \\
    (a) Congress & (b) Tags
\end{tabular}
\vspace{-4mm}
\caption{Effectiveness of intermediate node representations.}\label{fig:tfe}
\vspace{-5mm}
\end{figure}

\vspace{1mm}
\noindent
\textbf{Q3. Intermediate node representation.}
Figure~\ref{fig:tfe} shows the relative accuracy of {\method} according to how much the intermediate node embeddings are used in the inter-snapshot learning.
The \textit{more} intermediate node embeddings are used in the temporal feature update module of {\method}, 
{\method} could achieve the \textit{higher} hyperedge prediction accuracy.
These results imply that the intermediate node embeddings can provide useful information in capturing the temporal patterns of high-order relations.

\section{Conclusion}\label{sec-con}
In this paper, 
we observe two important yet under-explored characteristics of real-world networks:
high-order relations tend to \textbf{(O1)} \textit{have a structural and temporal influence on others in a short term} and \textbf{(O2)} \textit{periodically re-appear in a long term}.
To address them, 
we propose {\method} that employs (1) bi-interactional hyperedge encoding for (O1) and (2) periodic time injection and (3) intermediate node representation (O2).
Via extensive experiments on seven real-world datasets,
we verify the superiority of {\method} in capturing the high-order dynamics of real-world networks.

\begin{acks}\label{sec:ack}
This work was supported by Institute of Information \& Communications Technology Planning \& Evaluation (IITP) grant funded by the Korea government (MSIT) (RS-2022-00155586, 2022-0-00352).
\end{acks}

\clearpage
\section{GenAI Usage Disclosure}\label{sec-genai}
In accordance with the ACM authorship policy, we disclose the usage of generative AI tools (e.g., ChatGPT) as follows.
\begin{itemize}[leftmargin=10pt]
    \item \textbf{GenAI usage in writing}: ChatGPT was only used to review grammatical consistency during the writing the manuscript.
    \item \textbf{GenAI usage in data processing}: ChatGPT was only used for generating code for drawing figures (e.g., matplotlib.pyplot).
    \item \textbf{Author responsibility}: All uses of GenAI were limited to assistant roles. We conducted final verification and refinement of all GenAI generated results, analyses, and textual content.
\end{itemize}


\bibliographystyle{ACM-Reference-Format}
\bibliography{reference}

\end{document}